\documentclass{vgtc}                          % final (conference style)
\ifpdf%                                % if we use pdflatex
  \pdfoutput=1\relax                   % create PDFs from pdfLaTeX
  \usepackage{graphicx}                % allow us to embed graphics files
  \DeclareGraphicsExtensions{.pdf,.png,.jpg,.jpeg} % for pdflatex we expect .pdf, .png, or .jpg files
\else%                                 % else we use pure latex
  \usepackage{graphicx}                % allow us to embed graphics files
  \DeclareGraphicsExtensions{.eps}     % for pure latex we expect eps files
\fi%

\graphicspath{{figures/}{pictures/}{images/}{./}} % where to search for the images

\usepackage{microtype}                 % use micro-typography (slightly more compact, better to read)
\PassOptionsToPackage{warn}{textcomp}  % to address font issues with \textrightarrow
\usepackage{textcomp}                  % use better special symbols
\usepackage{mathptmx}                  % use matching math font
\usepackage{times}                     % we use Times as the main font
\usepackage{cite}                      % needed to automatically sort the references
\usepackage{tabu}                      % only used for the table example
\usepackage{booktabs}                  % only used for the table example
\usepackage{url}                       % to show URL in references
\usepackage{xcolor}

\onlineid{0}

\vgtccategory{Research}

\vgtcinsertpkg

\title{An Enhanced MA Plot with R-Shiny \\to Ease Exploratory Analysis of Transcriptomic Data}

\author{Ali Sheharyar\thanks{Code available under GPL-3 at \url{https://github.com/alisheharyar/Enhanced_MA_Plot}  license.}\\%e-mail: ali.sheharyar@qatar.tamu.edu}\\ %Sheharyar, Ali <ali.sheharyar@qatar.tamu.edu>
        \scriptsize Texas A\&M University at Qatar%
\and Talar Boghos Yacoubian\\ %\thanks{e-mail: tbyacoubian@hbku.edu.qa}\\ %Talar Boghos Yacoubian <tbyacoubian@hbku.edu.qa>
     \scriptsize Hamad Bin Khalifa University, CSE%
\and Dina Aljogol\\ %\thanks{e-mail: daljogol@hbku.edu.qa}\\ %
      \scriptsize Hamad Bin Khalifa University, CHLS%
\and Borbala Mifsud\\ %\thanks{e-mail: bmifsud@hbku.edu.qa}\\ %
      \scriptsize Hamad Bin Khalifa University, CHLS%
\and Dena Al Thani\\ %\thanks{e-mail: dalthani@hbku.edu.qa}\\ %
      \scriptsize Hamad Bin Khalifa University, CSE%
\and Michael Aupetit\thanks{Contact: ali.sheharyar@qatar.tamu.edu, maupetit@hbku.edu.qa}\\ %
      \scriptsize Qatar Computing Research Institute, HBKU, Qatar%
     }

\abstract{MA plots are used  to analyze the genome-wide differences in gene expression between two distinct biological conditions. An MA plot is usually rendered as a static scatter plot. Our interview with 3 experts in genomics showed that we could improve the usability of this plot by adding interactive analytic features. In this work we present the design study of the enhanced MA plot.}
\CCScatlist{
  \CCScatTwelve{Human-centered computing}{Visu\-al\-iza\-tion}{Visu\-al\-iza\-tion techniques};
}

\begin{document}


\maketitle

%%%%%%%%%%%%%%%%%%%%%%%%%%%%%%%%%%%%%%%%%%
%%%%%%%%%%%%%%%%%%%%%%%%%%%%%%%%%%%%%%%%%%
\section{Introduction} %for journal use above \firstsection{..} instead
\label{sec:intro}
An MA plot is used in biomedical research to visualize differential gene expression between two conditions e.g. control vs treated samples, two cell types etc. The difference could be measured either by microarrays or RNA-sequencing, where the gene expression level of each gene in every sample is determined by the probe signal intensity (microarrays) or read counts (RNA-sequencing) in the corresponding experiment (\cite{Yang2002}, \cite{Wang2010}). For example, the read count of gene X indicates how much of gene X is produced in that sample.

The MA plot is a scatter plot, and each point represents a single gene. The x-axis shows the average expression level of the gene across the two conditions (A=$0.5 \times \log_{2}(RG)$). The y-axis is the $\log_{2}$ fold change between the samples (M=$\log_{2}(R/G)$), where R and G are the signal intensities or read counts of a given gene in the two conditions. Those with similar expression levels in both conditions will be closer to y = 0. A statistical test needs to be performed to identify significantly differential genes. A p-value threshold is set \textit{a priori} to exclude potentially random events that may arise due to variation in biological and technical replicates. Significantly different genes are colored red, when the difference is positive and blue if it is negative. Genes with no significant differences are colored in grey, and those with missing p-values are indicated in yellow (see Figure \ref{fig:snapshot}). 

%%%%%%%%%%%%%%%%%%%%%%%%%%%%%%%%%%%%%%%%%%
%%%%%%%%%%%%%%%%%%%%%%%%%%%%%%%%%%%%%%%%%%
\section{Users and their Needs}

In the course of designing tools to support experts in RNA-seq data analysis, we interviewed 3 domain experts to get feedback on typical visualizations they used, among which the MA plot, focus of this work. 

These 3 experts are: A) a Ph.D. student in Genomics and Precision Medicine and a holder of M.Sc. in Molecular Cell Biology; B) an Assistant Professor in Health and Life Sciences, a PhD in Molecular Biology and a frequent user of MA plots in regular research; and C) a PhD in Genetics currently working as a Principal Investigator/Director of a Research Team with heavy experience in RNA-Seq research and corresponding MA data. 
All three users are highly specialized in the area of focus of this study and are well-versed in complex statistical modelling in a mix of programming languages and corresponding analytical environments. 

They expressed the needs in the next section, that were not satisfied in existing tools displaying MA plots.

%%%%%%%%%%%%%%%%%%%%%%%%%%%%%%%%%%%%%%%%%%
%%%%%%%%%%%%%%%%%%%%%%%%%%%%%%%%%%%%%%%%%%
\section{Design Goals and Technical Solutions}
\label{sec:design}
\begin{figure*}[h!]
 \centering 
 \fbox{\includegraphics[width=0.9\linewidth]{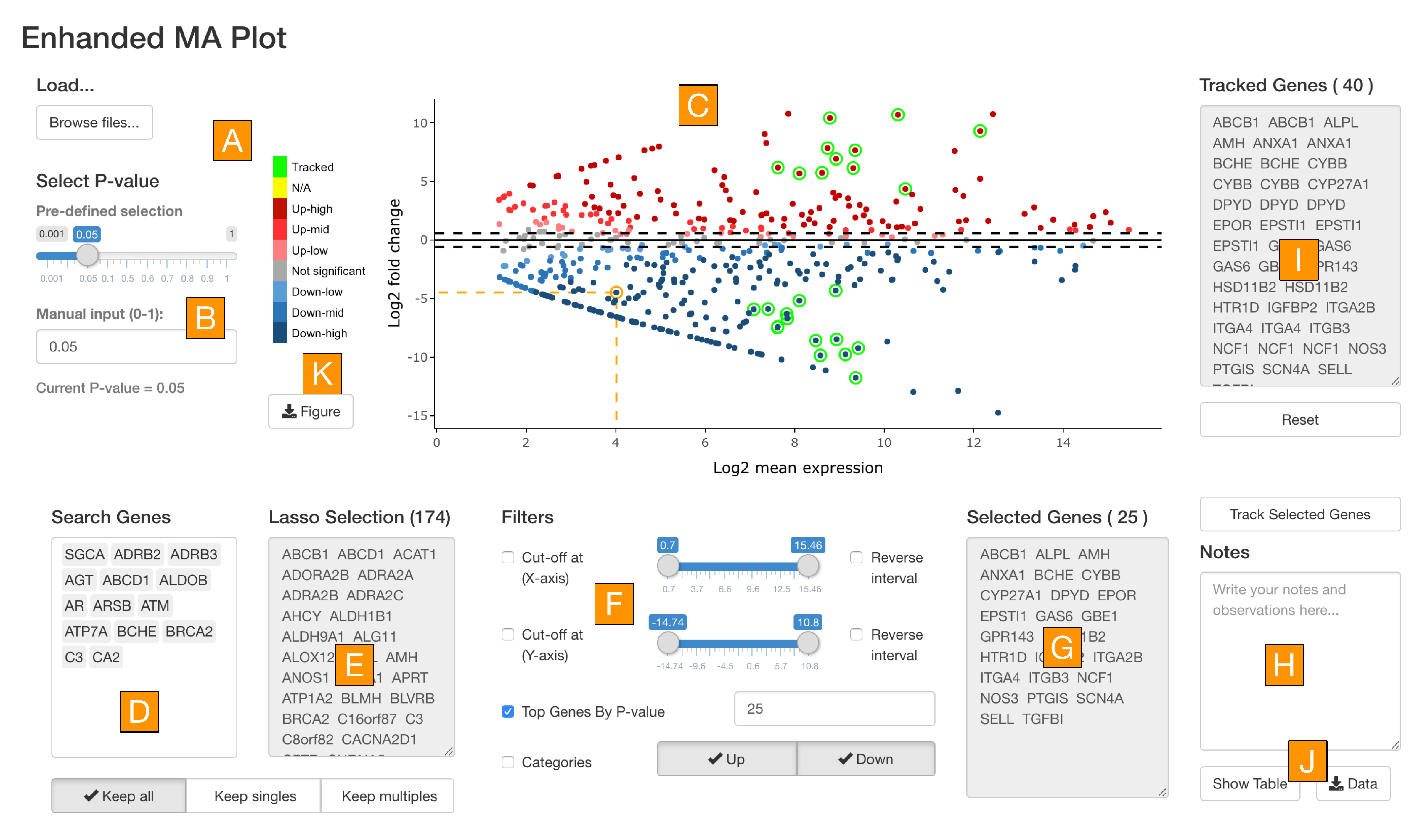}}
 \caption{Enhanced MA plot interface. See description in Section \ref{sec:design}.}
 \label{fig:snapshot}
\end{figure*}

All the expressed needs below required an \textit{interactive} MA plot and additional widgets (Capital letters below refer to the illustration in Figure \ref{fig:snapshot}), and became our \textbf{design goals}. We give the technical solutions for each of them following guidelines from Munzer \cite{Munzner2014}. 

\noindent\textbf{Load data from a file and Display the MA plot}
The user can load data from a \texttt{CSV} file (A) and display them as a standard MA plot visualization (C) as described in Section \ref{sec:intro}.

\noindent\textbf{Change interactively the significance level and enable pan and zoom}
The user can use a slider to select the level of significance P with predefined typical values (0.01, 0.05, 0.1...) or enter an exact value in a text area (B). Pan and zoom are native interactions in the R-Plotly \cite{plotly} scatter plot widget (C).

\noindent\textbf{Get the name of a gene in the plot}
The user gets details like gene name, M, A and P values of the gene hovered with the mouse pointer in the scatter plot (C).

\noindent\textbf{Select genes graphically}
The user can select genes in the scatter plot using a lasso and a box selection (C,E).

\noindent\textbf{Search and select a gene by name}
The user can search a particular genes by entering in a text area (D) its name. As the user types, the text area will start searching from the list of all genes that partially match the input string. The matching genes are displayed in a list to choose from. A mouse click on any selected gene highlights its location in the scatter plot (C).

\noindent\textbf{Combine all these selections with standard Boolean operations} 
The user can get the union (\textit{Keep all}), the intersection (\textit{Keep multiples}) or the mutually exclusive (\textit{Keep singles}) genes from the lasso/box and search selected sets (D,E).

\noindent\textbf{Filter specific genes by largest significance level, and by M and A values}
The user can filter the genes with largest p-value, keeping the top K. The user can filter genes by range on M and A setting values of $M_{\min}$, $M_{\max}$, $A_{\min}$ and $A_{\max}$ with a check box input to select the inside or outside range (F).

\noindent\textbf{Track the selected genes} 
The user can track the genes resulting from the filter (G), highlighting them (green outline) in the scatter plot by pushing the "Track Selected Genes" button (I). 

\noindent\textbf{Export the figures and the selected genes for further analysis with other plots or tools}
The user can download (J) the resulting genes as a \texttt{csv} file, and all R objects and ggplot as an \texttt{Rdata} file, and also download (K) the MA plot as a \texttt{png} file.

%%%%%%%%%%%%%%%%%%%%%%%%%%%%%%%%%%%%%%%%%%

%%%%%%%%%%%%%%%%%%%%%%%%%%%%%%%%%%%%%%%%%%
%%%%%%%%%%%%%%%%%%%%%%%%%%%%%%%%%%%%%%%%%%
\section{Layout Design of the Interface}
\label{sec:layout}

Beyond the expressed needs, when observing the usage of an earlier prototype, \textbf{users tend to lasso the most significant genes, away from the grey dots in the medium area, for further analysis}. Hence we use a diverging color map (A) that displays the most significant genes in darker shades of red and blue, and thus eases their lasso selection.
We also noticed the user were speaking out loud reasons for their choice, hence we added a take-note text area (H) that is saved in the \texttt{Rdata} file.

\begin{figure}[h!]
 \centering % avoid the use of \begin{center}...\end{center} and use \centering instead (more compact)
 \fbox{\includegraphics[width=0.8\columnwidth]{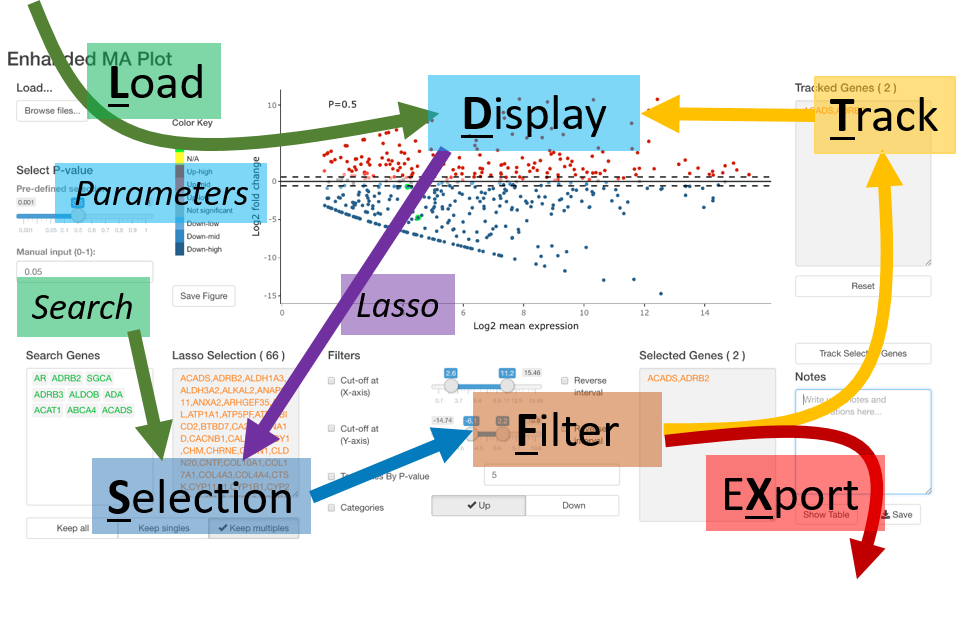}}
 \caption{Design of the interface to support the cycling \textbf{L-DSFT-DSFT-...-DSFT-X} interaction flow. See Section \ref{sec:layout}.}
 \label{fig:cycle}
\end{figure}

We observed that when genes were tracked (highlighted), the users wanted to use selection and filter again either to refine the tracked set of genes using other filters, or to expand the tracked set by lassoing neighboring genes in the MA plot.
Hence we designed the layout to ease that \textit{Display-Select-Filter-Track} (\textbf{DSFT}) sequential analytic process (See Figure \ref{fig:cycle}). Following standard Western reading "Z" pattern, the initial data Loading step (\textbf{L} (A)) had to be put on the top left of the interface, while the output result of the analytic process, eXporting (\textbf{X} (J)) selecting genes, had to be put on the  bottom right. So we organize the interface to cycle counter-clockwise through the \textbf{L$-$DSFT$-$DSFT$-$...$-$DSFT$-$X} steps (\textbf{DSFT} (C,D,E,F,G,I)). 

%%%%%%%%%%%%%%%%%%%%%%%%%%%%%%%%%%%%%%%%%%
%%%%%%%%%%%%%%%%%%%%%%%%%%%%%%%%%%%%%%%%%%
\section{Summative Evaluation}
We ran a summative evaluation regarding the usability of the resulting interface with two of the users (A and B). To qualitatively assess user experience, we used a think-aloud protocol followed by the System Usability Scale (SUS) \cite{sus}, a tool to measure and quantify the perception of usability.

The overall SUS score of user A is 95, corresponding to an adjective rating of `Excellent', and that of user B is 65, corresponding to an adjective rating of `OK' or `Fair'.  These results reflect findings relating to learnability, usability, and Net Promoter Score which reflects the likelihood of recommending the tool to other users.
The user think-aloud protocol feedback has highlighted the user needs of understanding details of the datasets used and being informed of the format, preprocessing and loading requirements of the tool. Both users expressed that the current version enhanced the previous prototypes and that they are willing to use the tool and test it on datasets for their current research. They confirmed the usefulness of the tool for exploratory analysis. 

%%%%%%%%%%%%%%%%%%%%%%%%%%%%%%%%%%%%%%%%%%
\vspace{-0.5cm}
%%%%%%%%%%%%%%%%%%%%%%%%%%%%%%%%%%%%%%%%%%
%%%%%%%%%%%%%%%%%%%%%%%%%%%%%%%%%%%%%%%%%%
\section{Technical Details}
We developed this tool with the \texttt{Shiny} \cite{Rshiny} R framework using \texttt{Plotly} \cite{plotly} package to add interaction to the MA scatter plot. 
The MA scatter plot is produced using the modified version of the \texttt{ggmaplot} function from the \texttt{ggpubr} \cite{ggpubr} R package.

The code is available at \url{https://github.com/alisheharyar/Enhanced_MA_Plot} under GPL-3 license. It can be tested online at: \url{https://michaelaupetit.shinyapps.io/EnhancedMAplot_V2/} 

\vspace{-0.2cm}
%%%%%%%%%%%%%%%%%%%%%%%%%%%%%%%%%%%%%%%%%%
%%%%%%%%%%%%%%%%%%%%%%%%%%%%%%%%%%%%%%%%%%
\section{Conclusion}
We proposed an enhanced MA plot with added interactive features that could improve the visualization and analysis of differentially expressed genes. We performed a summative assessment to evaluate its usability and usefulness. The domain experts appreciated the tool and considered it an improvement to the state-of-the-art. They both stated that they would like to use it for exploratory analysis purposes. 

% The authors wish to thank A, B, and C. This work was supported in part by
% a grant from XYZ.}

%\bibliographystyle{abbrv}
\bibliographystyle{abbrv-doi}

\bibliography{template}

\begin{thebibliography}{1}

\bibitem{sus}
A.~S. f.~P. Affairs.
\newblock System usability scale (sus), Sep 2013.

\bibitem{Rshiny}
W.~Chang.
\newblock {\em {shiny}: Web Application Framework for {R}}, 2015.
\newblock RStudio, package version 1.5.0,
  \url{https://CRAN.R-project.org/package=shiny}.

\bibitem{ggpubr}
A.~Kassambara.
\newblock {\em ggpubr: 'ggplot2' Based Publication Ready Plots}, 2020.
\newblock R package version 0.2.5.

\bibitem{Munzner2014}
T.~Munzner.
\newblock {\em Visualization Analysis and Design}.
\newblock {A.K.} Peters visualization series. A {K} Peters, 2014.

\bibitem{plotly}
C.~Sievert.
\newblock {\em Interactive Web-Based Data Visualization with R, plotly, and
  shiny}.
\newblock Chapman and Hall/CRC, 2020.

\bibitem{Wang2010}
L.~Wang, Z.~Feng, X.~Wang, X.~Wang, and X.~Zhang.
\newblock Degseq: an r package for identifying differentially expressed genes
  from rna-seq data.
\newblock {\em Bioinformatics}, 26(1):136--138, 2010.

\bibitem{Yang2002}
Y.~H. Yang, S.~Dudoit, P.~Luu, D.~M. Lin, V.~Peng, J.~Ngai, and T.~P. Speed.
\newblock Normalization for cdna microarray data: a robust composite method
  addressing single and multiple slide systematic variation.
\newblock {\em Nucleic acids research}, 30(4):e15--e15, 2002.

\end{thebibliography}
\end{document}